\documentclass[twocolumn,prl, aps,letterpaper,groupedaddress]{revtex4-1}

\usepackage{times,amsmath,amsfonts,amssymb,latexsym,textcomp}
\usepackage{color}
\usepackage{graphicx}
\usepackage{multirow}
\usepackage{bbm}
\usepackage{tabularx}
\usepackage{amsthm}
\usepackage{dsfont}
\usepackage{float}
\usepackage{array}
\usepackage{braket}

\newcommand\id{\ensuremath{\mathbbm{1}}}

\begin{document}
\title{Linear-optical generation of eigenstates of the two-site XY model}
\author{Stefanie Barz$^{1,\dagger}$, Borivoje Daki{\'c}$^{1,2}$, Yannick~Ole~Lipp$^1$,   Frank Verstraete$^1$, James D. Whitfield$^1$, Philip Walther$^1$}
\affiliation{$^1$~University of Vienna, Faculty of Physics, Boltzmanngasse 5, 1090 Vienna, Austria\\
$^2$~Institute for Quantum Optics and Quantum Information (IQOQI), Austrian Academy of Sciences, Boltzmanngasse 3, A-1090 Vienna, Austria\\
$^\dagger$~Present address: Clarendon Laboratory, Department of Physics, University of Oxford, Parks Road, Oxford OX1 3PU, UK}

\begin{abstract}
Much of the anticipation accompanying the development of a quantum computer relates to its application
to simulating dynamics of another quantum system of interest. Here we study the building blocks
for simulating quantum spin systems with linear optics. We experimentally generate the eigenstates of the XY Hamiltonian under an external magnetic field. The implemented quantum circuit consists of two CNOT gates, which are realized experimentally by harnessing entanglement from a photon source and by applying a CPhase gate. We tune the ratio of coupling constants and magnetic field by changing local parameters. This implementation of the XY model using linear quantum optics might open the door to the future studies of quenching dynamics using linear optics.
\end{abstract}
\maketitle

\section{Introduction}
In 1982, Richard Feynman proposed the idea for the efficient simulation of quantum systems~\cite{feyquantumsimulators}. Complex systems, whose properties cannot easily be computed with classical computers, can be simulated by other well-controllable quantum systems. In this way, an easily accessible system can be used for reproducing the dynamics and the quantum state of another system of study. The insight of having one controllable quantum system simulate another is what forms the foundation of quantum simulation. There are two different approaches for simulating quantum systems that have been implemented experimentally: analog and digital simulation.
Analog quantum simulators are designed to mimic a quantum system by reproducing its evolution in a faithful manner~\cite{qsreview, pqsreview}. Alternatively, the effect of the unitary evolution of a quantum system may be regarded as that of a quantum circuit acting on some initial state. This inspires the approach of a digital quantum simulator where the state of the system is encoded into qubits and processed via quantum logic gates~\cite{lloyduqs, abrcircuitmodel, guzmolen}. The main challenge --~apart from providing a sufficiently powerful quantum computer~-- lies in finding a way to decompose the Hamiltonian into a suitable form. 
Experimentally, basic quantum simulations of both types have been demonstrated as proof-of-concept experiments on several quantum architectures including trapped ions~\cite{friedenauerions, lanunidqsions, brittonions, kimions}, optical lattices~\cite{trotzkyol, simonol}, nuclear magnetic resonance~\cite{somaroonmr, negrevergnenmr, brownnmr, dunmr, Peng2010} and photons~\cite{kassalph, mafrust, Orieux2013}.
 
Here we exploit a scalable approach for digital quantum simulation for strongly interacting Hamiltonians, which has been suggested in~\cite{Verstraete2008}. 
The general idea of~\cite{Verstraete2008} is to construct the explicit finite quantum circuits that transform the Hamiltonian into one corresponding to non-interacting particles.
In this work, we apply this method to the XY Hamiltonian for two spins in a magnetic field. 
We develop a quantum circuit that transforms product input states to the eigenstates of this Hamiltonian. 
Our approach allows us to recover the whole spectrum of certain quantum many-body problems - a distinct advantage of our implementation.
We experimentally implement this circuit in a linear optical setup and generate ground and excited states for the two-qubit Heisenberg XY model in a transverse external field. Our circuit consists of two CNOT gates, where one of the gates is absorbed in the state generation and this other is implemented physically.

\section{Theory}
\label{sec:theory}
Our work focuses on the simulation of the two-qubit XY Hamiltonian in a transverse external field:
\begin{equation}
	H = J_x \; \sigma_x \otimes \sigma_x + J_y \; \sigma_y \otimes \sigma_y + \frac{1}{2} B \left( \sigma_z \otimes \id + \id \otimes \sigma_z \right) \, ,
	\label{eqn:xyhamiltonian}
\end{equation}
with $J_x$ and $J_y$ being coupling constants, $B$ the magnetic field with unit magnetic moment, and $\sigma_{i}$ the Pauli matrices that represent the particles' spin in $x, y$ or $z$ direction, respectively. 
\begin{figure}
  \centering
	\includegraphics[width=.45\textwidth]{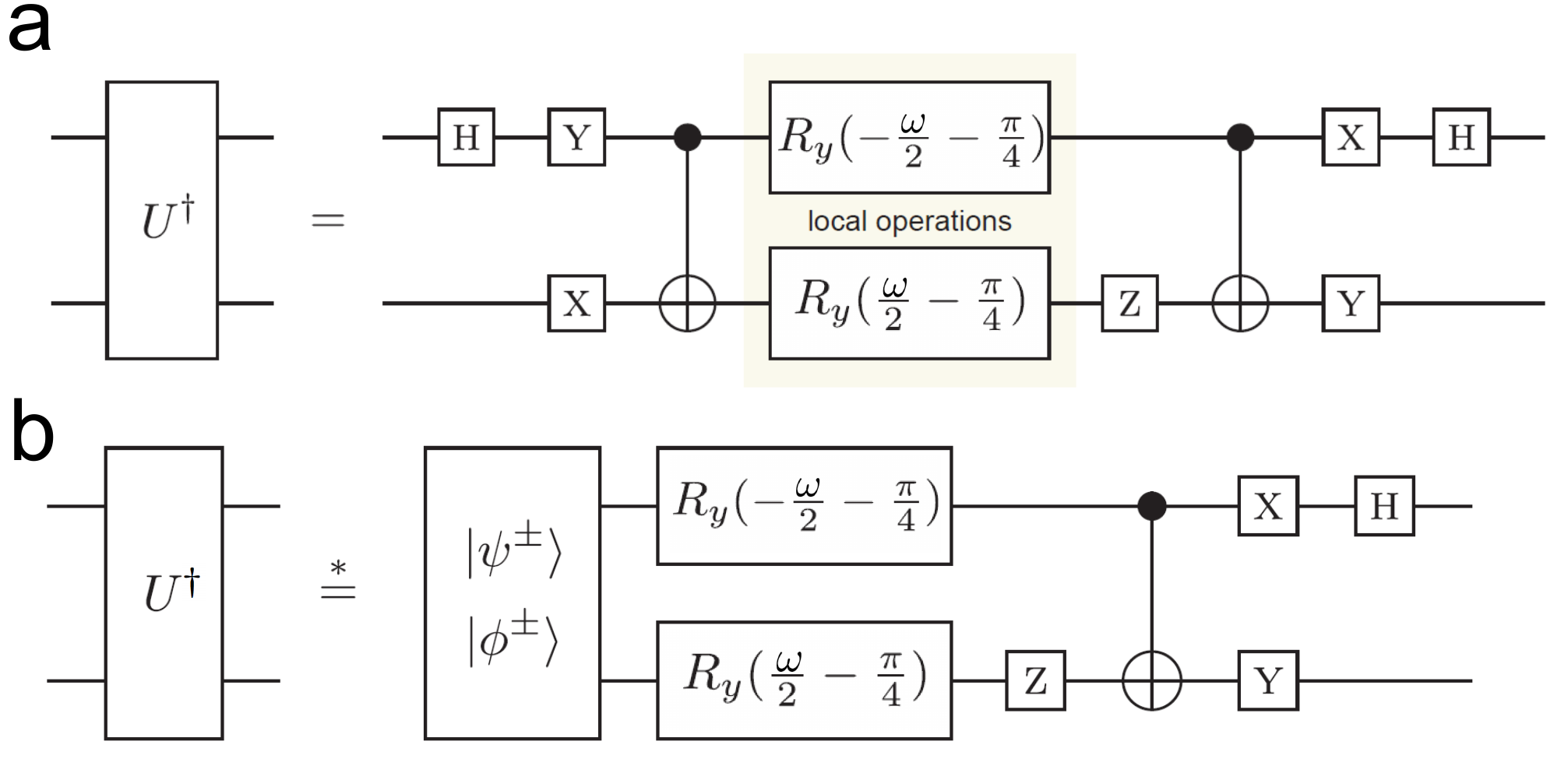}
	\caption{\label{Figure1}The quantum circuit for the generation of the eigenstates of the XY Hamiltonian. \textbf{a} The decomposition of the unitary $U^\dag$ that transforms the eigenstates of a non-interacting Hamiltonian to the eigenstates of the XY Hamiltonian. Full control of the system parameter $w$ that specifies a particular Hamiltonian is granted by local operations. 
	\textbf{b} The local unitaries $(YH)\otimes X$ together with the first CNOT transform the computational basis states into the four Bell states and we obtain the circuit shown in (b). 	The asterisk (*) is to remind us that this simplification is valid only for certain inputs.}
\end{figure}

Our work here focuses on preparing eigenstates. Specifically, our goal is to find a unitary $U$ that transforms the Hamiltonian, $H$, into one corresponding to non-interacting quasi-particles, $\tilde{H}$, hence diagonalizing it as:
\begin{equation}
\tilde{H}=UHU^{\dagger}=\omega_1\sigma_z\otimes\openone+\omega_2\openone\otimes\sigma_z,
\label{eq:diag}
\end{equation}
where $\omega_1=\frac{E_1+E_2}{2}$ 
and
$\omega_2=\frac{E_1-E_2}{2}$  
are the quasi-particle energies, $\tan{w}= (J_x-J_y)/B$, and $E_1$, $E_2$ are the eigenenergies. 
It is not difficult to verify that the desired unitary U is given by:
\begin{eqnarray}
U&=&\left(
    \begin{array}{cccc}
      \cos\frac{w}{2} & 0 & 0 & \sin\frac{w}{2} \\
      0 & \frac{1}{\sqrt2} & \frac{1}{\sqrt2} & 0 \\
      0 & -\frac{1}{\sqrt2} & \frac{1}{\sqrt2} & 0 \\
      -\sin\frac{w}{2} & 0 & 0 & \cos\frac{w}{2} \\
    \end{array}
  \right), 
\end{eqnarray}
and that \mbox{$E_1= \sqrt{B^2 + (J_x-J_y)^2}=-E_4$}, \mbox{$E_2= J_x + J_y=-E_3$}.

By applying $U^\dagger$ to the computational basis states, the eigenstates of $\tilde{H}$, we obtain the eigenstates of the Hamiltonian $H$:   
\begin{align}
	\ket{\psi_1} &= U^\dagger \ket{00} = \cos \frac{w}{2} \ket{00} + \sin \frac{w}{2} \ket{11} \\
	\ket{\psi_2} &= U^\dagger \ket{01} = \frac{1}{\sqrt{2}} \, (\ket{01} + \ket{10})\\
	\ket{\psi_3} &= U^\dagger \ket{10} = \frac{1}{\sqrt{2}} \, (\ket{01} - \ket{10}) \\
	\ket{\psi_4} &= U^\dagger \ket{11} = -\sin \frac{w}{2} \ket{00} + \cos \frac{w}{2} \ket{11}\, .
\end{align}

We directly constructed the circuit but to generalize to arbitrary length XY spin chain the more general case was presented in~Ref.~\cite{Verstraete2008}.  The steps for the general case are 1) identify spins with fermionic modes, 2) Fourier transform the fermionic modes, and 3) perform a Bogoliubov transformation to diagonalize the free fermions.  The first step requires only relabeling while the second and third steps require actual transformations.  The Fourier transform over $L$ sites can be done in $L\log L$ steps. An additional $L^2$ gates are needed to account for the antisymmetry in the fermionic basis~\cite{Verstraete2008}.  Finally, the Bogoliubov transformation requires only mixing positive and negative momenta pairwise.  The parameters of this mixing depend on $J_x,J_y,B$ and the momenta of the two Fourier modes being mixed.  The full Bogoliubov transformation can be done with $L/2$ gates which can all be done in parallel. The combined procedure implements a diagonalizing unitary similar to \eqref{eq:diag} in polynomial cost in $L$. 
Using photonic systems, the implementation of a heralded entangling gate requires two additional ancilla photons per two-qubit gate.

\section*{Experiment and Results} 
\label{sec:experiment}

\begin{figure}
	\centering
	\includegraphics[width=.85\columnwidth]{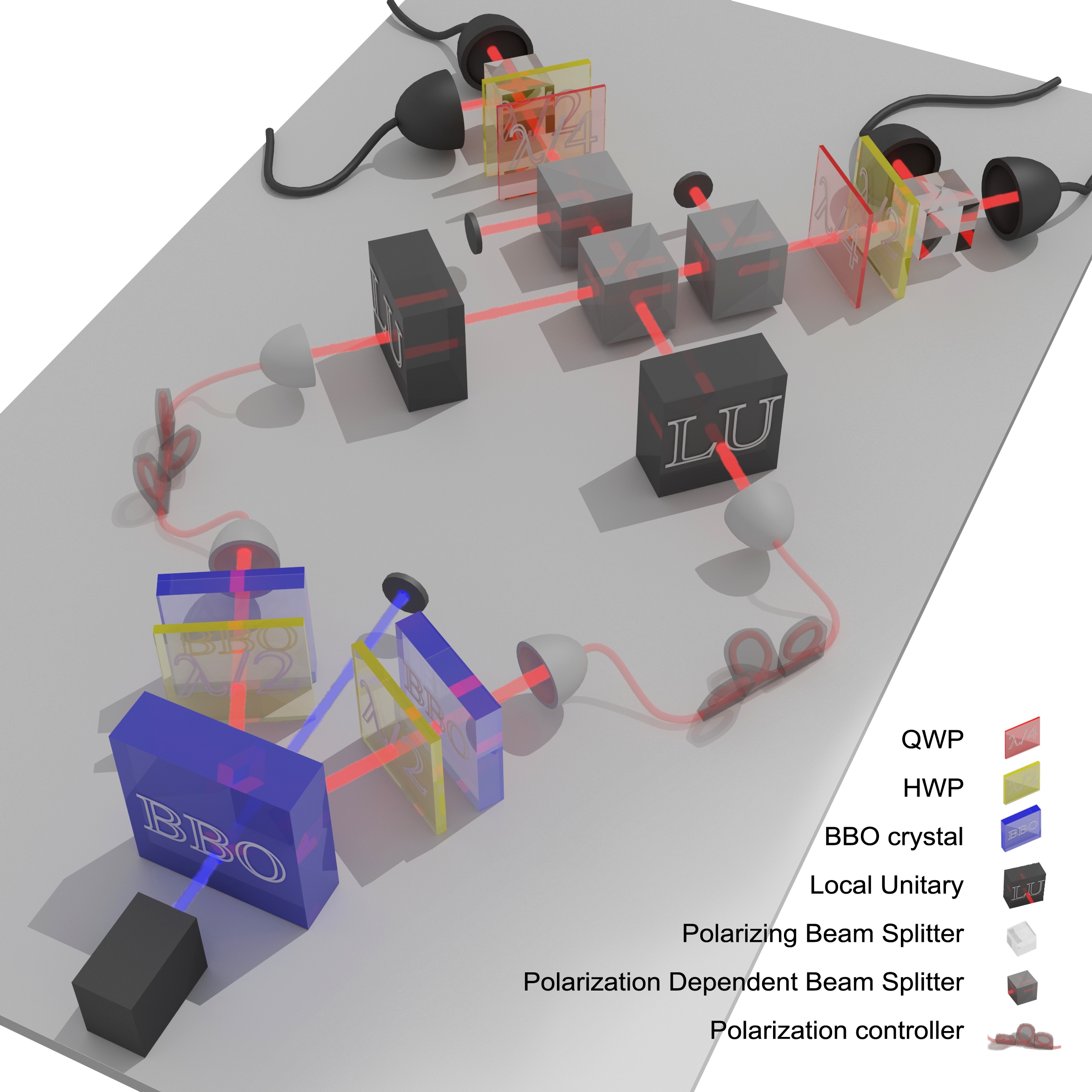}
	\caption{\label{Figure2}Schematic of the experimental setup implementing $U^\dagger$ (Fig.~\ref{Figure1}b). A pair of photons---polarization-encoded qubits---is both created and initialized to a desired state via spontaneous parametric down-conversion. Different Hamiltonians can flexibly be simulated be tuning the parameter $w=\arctan((J_x-J_y) / B)$ through local unitary operations (LU) wrapped in between two \textsc{cnot} operations. The former CNOT is thus merged with the SPDC process, while the latter is implemented by a combination of two-photon interference and polarization dependent beam splitters~\cite{lancsign, kiecsign, okacsign}.}
	\end{figure}

In our experiment, we generate the eigenstates using linear optics. To this end, we have experimentally realized a flexible optical circuit (Fig.~\ref{Figure2}) that may implement any unitary $U^\dagger(w)$ corresponding to the two-qubit XY Hamiltonian~\eqref{eqn:xyhamiltonian} in a transverse field with the system parameter $w$. Our circuit consists of two CNOT gates and local operations, which allow the manipulation of $w$.  Fig.~\ref{Figure1}a shows that the first CNOT gate together with the preceding unitaries in a transforms the four product state inputs $\ket{00}, \ket{01}, \ket{10}, \ket{11}$ into one of the Bell states $\ket{\psi^\pm}=(\ket{01}\pm\ket{01})/\sqrt{2}$, $\ket{\phi^\pm}=(\ket{00}\pm\ket{11})/\sqrt{2}$.
These Bell states can be naturally obtained by exploiting the entanglement of a spontaneous parametric down-conversion (SPDC) source. 
Thus, we integrate the first CNOT into the state preparation process (see Fig.~\ref{Figure1}b).
The input register in Fig.~\ref{Figure1}b originates from a type-II SPDC source, where a $\beta$-barium borate (BBO) crystal is pumped with a femtosecond-pulsed laser (394.5nm, 76MHz) to emit pairs of correlated photons at a wavelength of 789nm~\cite{Kwiat1999}. In our implementation, $\ket{0}$ and $\ket{1}$ correspond to horizontal and vertical polarization, respectively.
In our experiment, we generate entangled photons pairs in the four different Bell states and input them in the subsequent circuit.
In combination with narrow-bandwidth filters of 3nm this procedure yields state fidelities for the input states of $97 \pm 1 \%$.

Adjusting the subsequent local operations using a set of quarter-wave and half-wave plates allows us to tune the system parameter $w$. 
\begin{figure*}
	\centering
	\includegraphics[width=.85\textwidth]{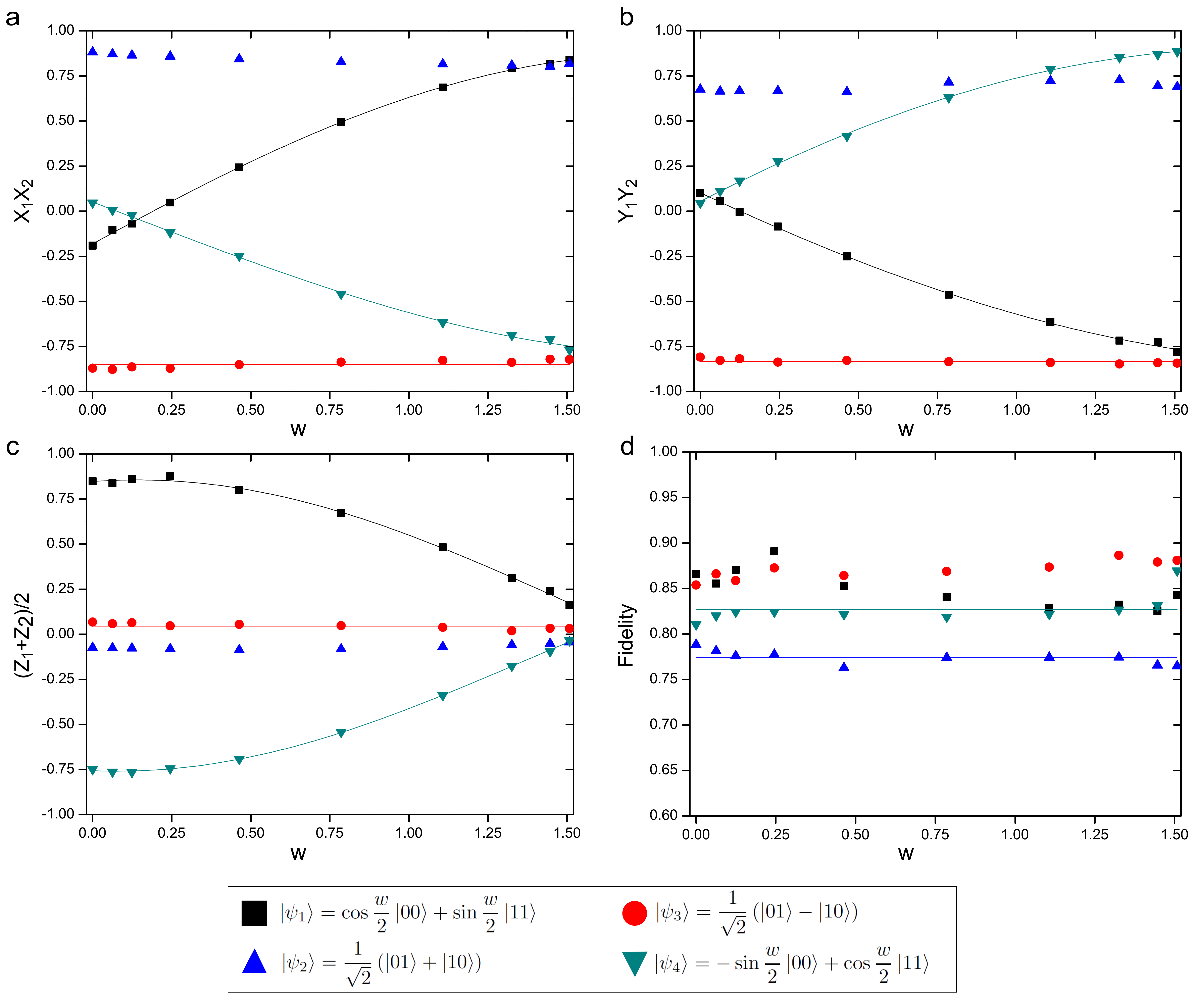}
	\caption{	\label{Figure3} Characterization of the eigenstates of the XY Hamiltonian. We show the measurement of the three components of the two-qubit XY Hamiltonian separately. \textbf{a} Correlations when measuring the qubits in the basis $\sigma_x\otimes\sigma_x$. \textbf{b} Correlations when measuring the qubits in the basis $\sigma_y\otimes\sigma_y$. \textbf{c} Correlations when measuring the qubits in the basis $(\sigma_z \otimes \id + \id \otimes \sigma_z)/2$.  This separate measurement of each part of the Hamiltonian makes it possible to see the measured data for each part separately and thus enables a more detailed analysis of each part of the Hamiltonian. \textbf{d} Fidelities of the generated states. In all subfigures, the lines show a fit of the experimental data.	
	The error bars are smaller than the point size and hence not shown.}
\end{figure*}

The circuit for $U^\dagger$ is completed by applying another CNOT gate. 
In our experiment, this destructive CNOT gate uses a polarization-dependent beam splitter (PDBS) which has a different transmission coefficient $T$ for horizontally polarized light ($T_H=1$) as for vertically polarized light ($T_V=1/3$)~\cite{Kiesel2005a}. 
If two vertically-polarized photons are reflected at this PDBS, they acquire a phase shift of $\pi$.
Two successive PDBSs with the opposite splitting ratios then equalize the output amplitudes.
This setup, in combination with two half-wave plates (HWPs) (see Figure~\ref{Figure2}) implements a destructive CNOT gate, where the success of the operation is determined by postselection on a coincidence detection of the final output photons~\cite{lancsign, Kiesel2005a, okacsign}. 
For this CNOT gate, we experimentally achieve a process fidelity~\cite{whimeasuringgates} of $86.0 \pm 0.3\%$.

Using this setup we are able to prepare both ground and excited states at arbitrary values of the system parameter $w$ by tuning the local unitaries in each input mode of the main polarization dependent beam-splitter.
Fig.~\ref{Figure3} shows different correlation measurements to characterize these states for several choices of $w$. 
In the Additional Information, we show the reconstructed density matrices of all measured states; Fig.~\ref{Figure3}d shows the state fidelities as obtained from the density matrices.

Our demonstration shows that the main features of the XY Hamiltonian can be reproduced.
The obtained fidelities lie between 0.75 and 0.9, these are the expected values when considering the fidelities of the entangled input states of 0.97 and a process fidelity of 0.86. Since the state fidelities of the experimental states are non-perfect, the measured data deviate from the theoretically expected values. However, as one can see in ~\ref{Figure3}, the obtained states show the same behavior as one would expect from the theoretical eigenstates.
In order to obtain data even closer to the values, one would need to increase the fidelity of the entangled input states, which are mainly limited by higher-order emissions in the current setup and can be increased using lower pump powers. Another limitation is the process fidelity of 0.86, which is mainly due to the non-perfect interference in our second CNOT gate. This interference could be improved by making the photons spectrally and spatially indistinguishable.
In summary, the current state fidelities are mainly limited due to technical challenges, which can be overcome.



\section{Conclusion} 
\label{sec:conclusion}
We have demonstrated the preparation of the eigenstates for the XY Hamiltonian under an external magnetic field.
In the original proposal~\cite{Verstraete2008}, it was pointed out that the same approach can also be applied to prepare thermal states and simulate the dynamical evolution of any integrable model. Other examples of integrable models are the Kitaev honey comb lattice~\cite{kitaevanalytical}, the 1-D Hubbard model and the Heisenberg models.

We end this paper with a discussion on the extension to dynamical studies. 
The importance of generating eigenstates is underlined in the context of quenching where the Hamiltonian of a system is instantaneously changed and the dynamics of a quantum system is examined. 
Recently, this problem has attracted significant interest~\cite{Polkovnikov2011,Cazalilla2010} and it is a difficult task to simulate the quantum dynamics classically.  In the quantum setting, utilizing algorithms such as the one implemented in the current work, initial states can be prepared and one could then perform evolution under a different Hamiltonian and observe the quenching properties for polynomial costs with a quantum computer. Since the XY model exhibits critical phases and quantum phase transitions, both adiabatic quenches through a phase transition and quenched dynamics can be studied using the present work as a starting point.
While we did not explore dynamics in the present work, future work might begin with the preparation of eigenstates and proceed to break integrability e.g.~by including an additional magnetic field in the $X$ direction and observing the dynamics of various observables. This will require the subsequent application of further entangling gates. However, currently the maximum number of subsequent photonic gates that has been demonstrated experimentally is two~\cite{lopez2012}, which can be increased to three~\cite{Barz2014} when using entangling input states as demonstrated here.

\section{Acknowledgments}
This work was supported from the European Commission, Q-ESSENCE (No. 248095), QUILMI (No. 295293), EQUAM (No. 323714), PICQUE (No. 608062), GRASP (No. 613024), and the ERA-Net CHISTERA project QUASAR, the John Templeton Foundation, the Ford Foundation, the Vienna Center for Quantum Science and Technology (VCQ), the Austrian Nano-initiative NAP Platon, the Austrian Science Fund (FWF) through the SFB FoQuS (F4006-N16), START (Y585-N20) and the doctoral programme CoQuS, the Vienna Science and Technology Fund (WWTF, grant ICT12-041), and the United States Air Force Office of Scientific Research (FA8655-11-1-3004).

%

\clearpage
\newpage
\onecolumngrid
\appendix

\section{\textbf{Additional information:} Analysis of the output states}
Here, we show the density matrices of the theoretical as well as the measured output states.

\begin{figure}[htbp]
	\centering
		\includegraphics[height=0.750\textheight]{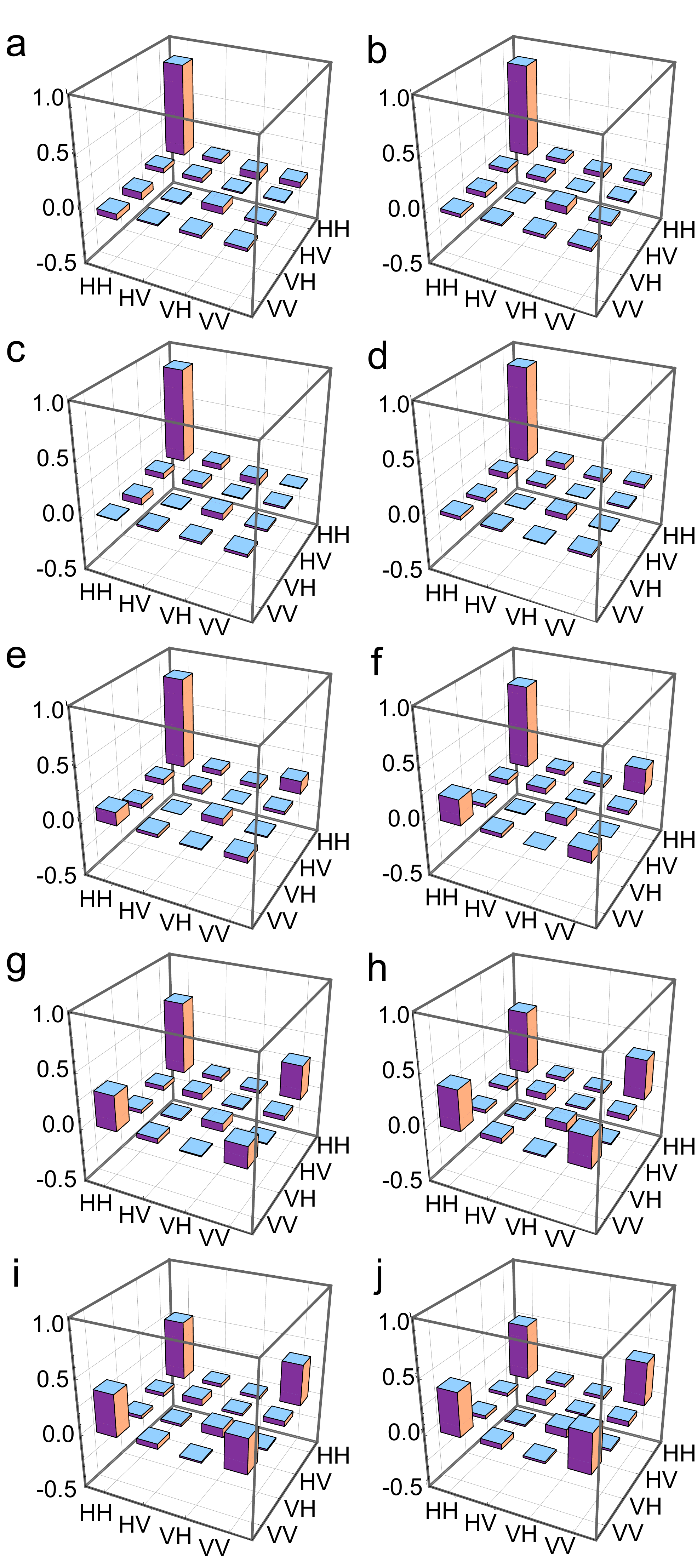}
	\label{fig:densitymatrices_phi+_1}
		\caption{Measured density matrices corresponding to the output state: $\ket{\psi_1}=\cos{(w/2)}\ket{00}+\sin{(w/2)}\ket{11}$ for different values of $w$ (real parts). a, $w=\arctan{(0)}$, b, $w=\arctan{(1/16)}$, c, $w=\arctan{(1/8)}$, d, $w=\arctan{(1/4)}$, e, $w=\arctan{(1/2)}$, f, $w=\arctan{(1)}$, g, $w=\arctan{(2)}$, h, $w=\arctan{(4)}$, i, $w=\arctan{(4)}$, j, $w=\arctan{(16)}$.}
\end{figure}

\begin{figure}[htbp]
	\centering
		\includegraphics[width=0.50\textwidth]{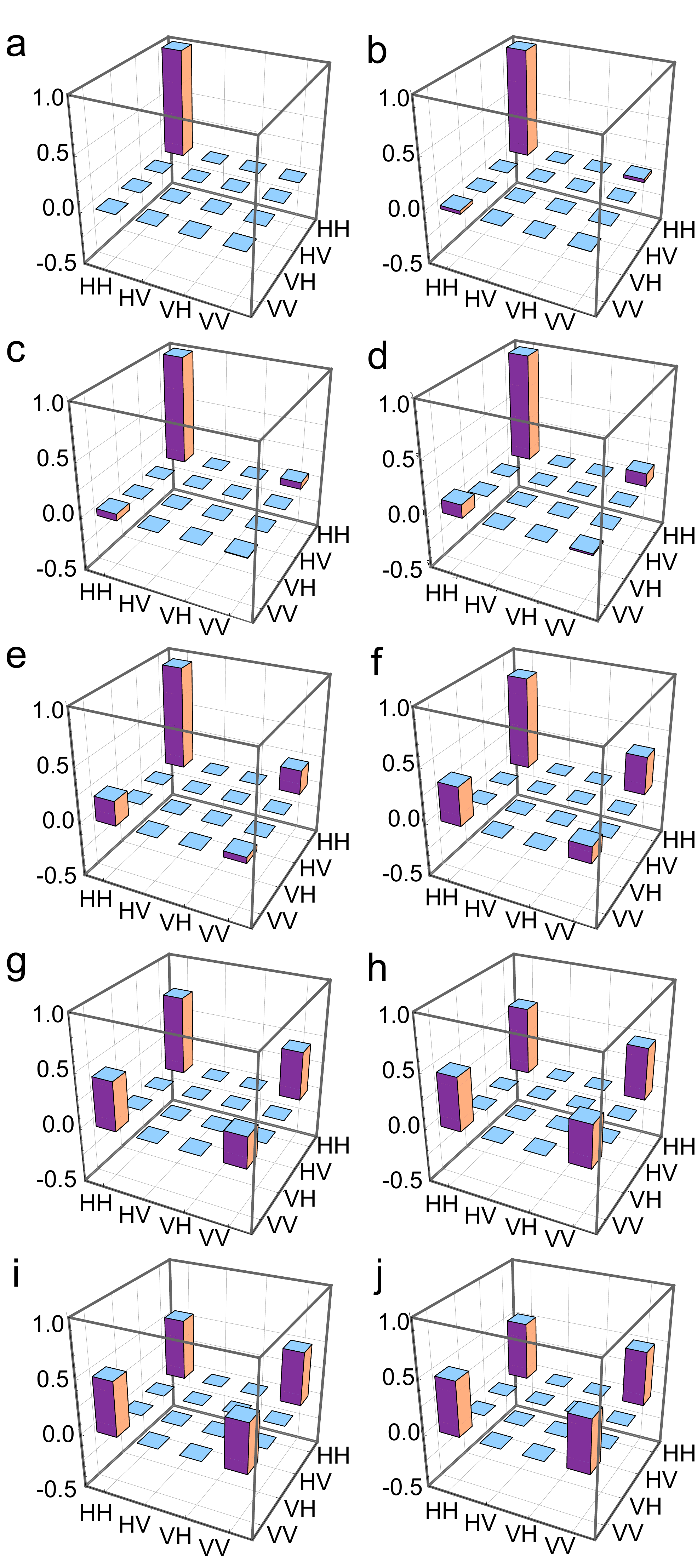}
		\label{fig:densitymatrices_phi+_1_theory}
		\caption{Theoretical density matrices corresponding to the output state: $\ket{\psi_1}=\cos{(w/2)}\ket{00}+\sin{(w/2)}\ket{11}$ for different values of $w$ (real parts). a, $w=\arctan{(0)}$, b, $w=\arctan{(1/16)}$, c, $w=\arctan{(1/8)}$, d, $w=\arctan{(1/4)}$, e, $w=\arctan{(1/2)}$, f, $w=\arctan{(1)}$, g, $w=\arctan{(2)}$, h, $w=\arctan{(4)}$, i, $w=\arctan{(4)}$, j, $w=\arctan{(16)}$.}
\end{figure}

\begin{figure}[htbp]
	\centering
		\includegraphics[width=0.50\textwidth]{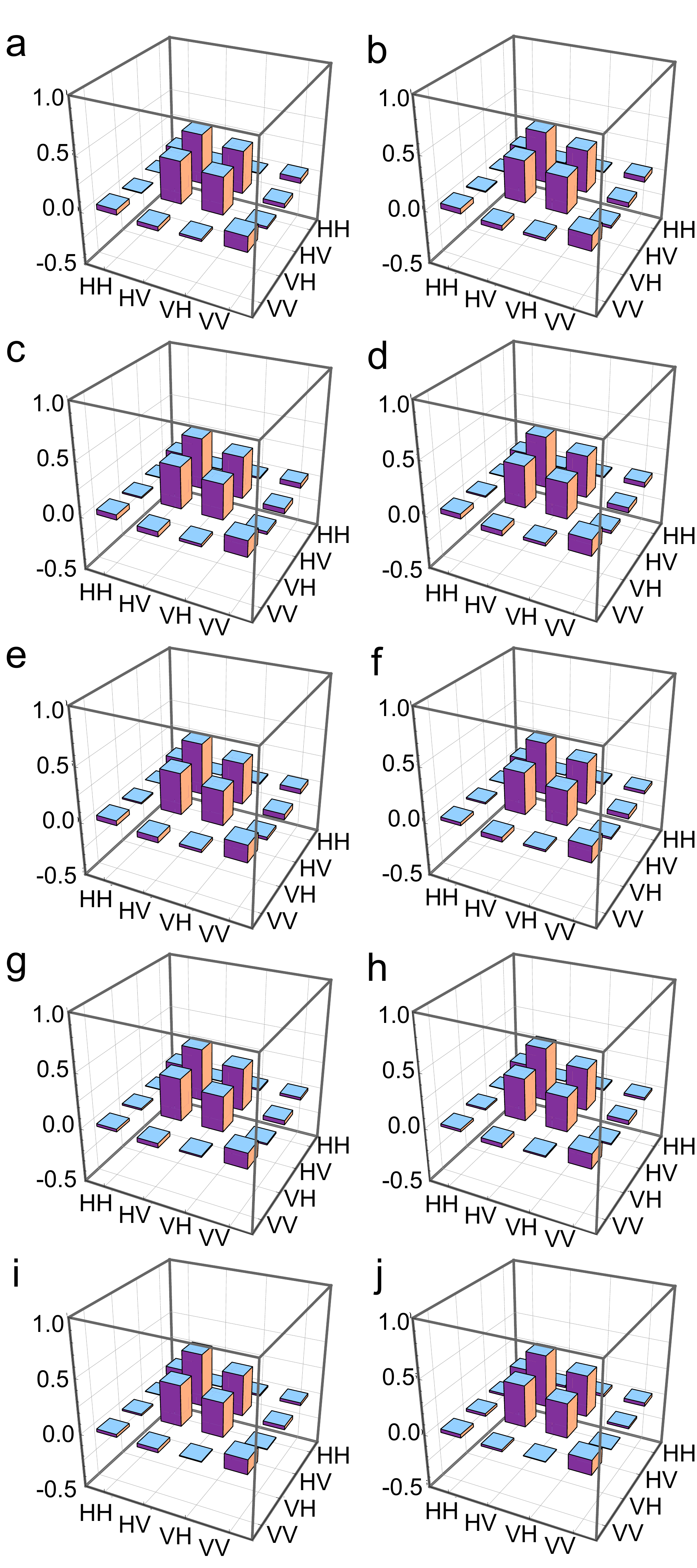}
			\label{fig:densitymatrices_psi+_3}
		\caption{Measured density matrices corresponding to the output state: $\ket{\psi_2}=(\ket{01}+\ket{10})/\sqrt{2}$ for different values of $w$ (real parts). a, $w=\arctan{(0)}$, b, $w=\arctan{(1/16)}$, c, $w=\arctan{(1/8)}$, d, $w=\arctan{(1/4)}$, e, $w=\arctan{(1/2)}$, f, $w=\arctan{(1)}$, g, $w=\arctan{(2)}$, h, $w=\arctan{(4)}$, i, $w=\arctan{(4)}$, j, $w=\arctan{(16)}$.}
\end{figure}

\begin{figure}[htbp]
	\centering
		\includegraphics[width=0.50\textwidth]{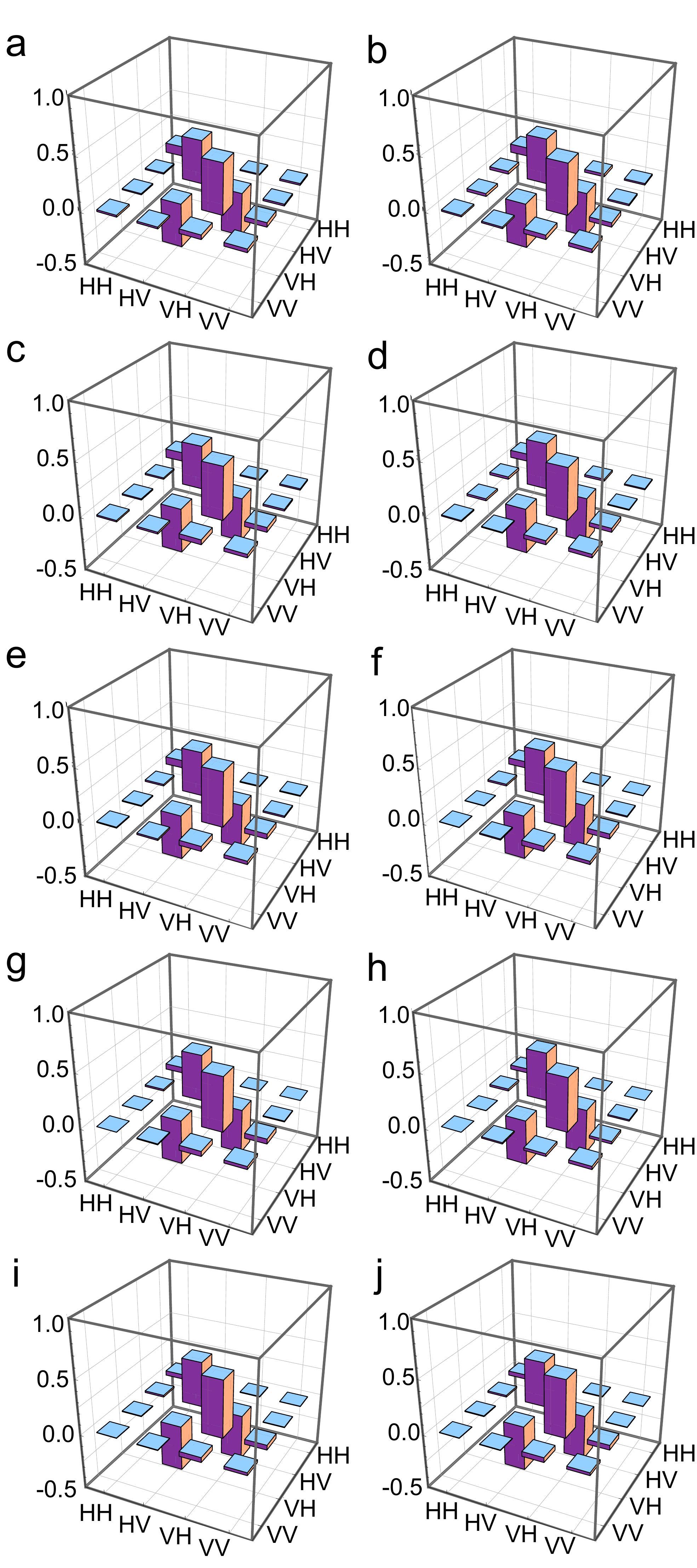}
	\label{fig:densitymatrices_phi-_2}
			\caption{Measured density matrices corresponding to the output state: $\ket{\psi_3}=(\ket{01}-\ket{10})/\sqrt{2}$ for different values of $w$ (real parts). a, $w=\arctan{(0)}$, b, $w=\arctan{(1/16)}$, c, $w=\arctan{(1/8)}$, d, $w=\arctan{(1/4)}$, e, $w=\arctan{(1/2)}$, f, $w=\arctan{(1)}$, g, $w=\arctan{(2)}$, h, $w=\arctan{(4)}$, i, $w=\arctan{(4)}$, j, $w=\arctan{(16)}$.}
\end{figure}

\begin{figure}[htbp]
	\centering
		\includegraphics[width=0.30\textwidth]{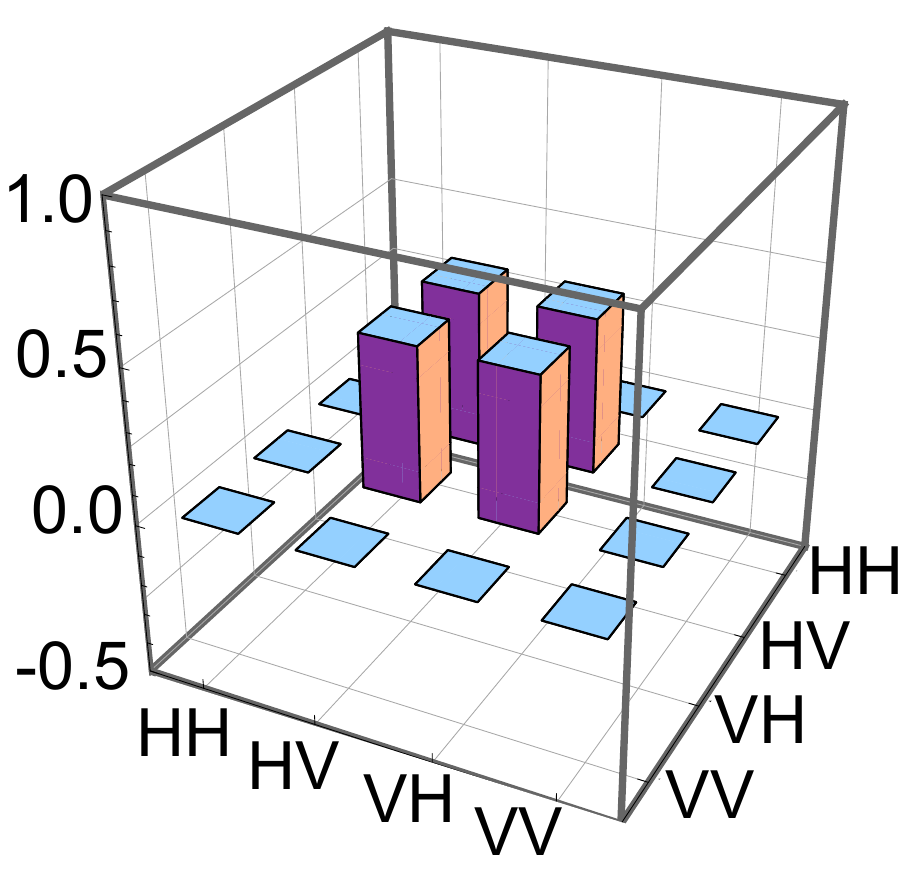}
	\label{fig:densitymatrices_psi+_3_theory}
			\caption{Theoretical density matrix corresponding to the output state: $\ket{\psi_2}=(\ket{01}+\ket{10})/\sqrt{2}$ for all values of $w$ (real parts).}
\end{figure}

\begin{figure}[htbp]
	\centering
		\includegraphics[width=0.30\textwidth]{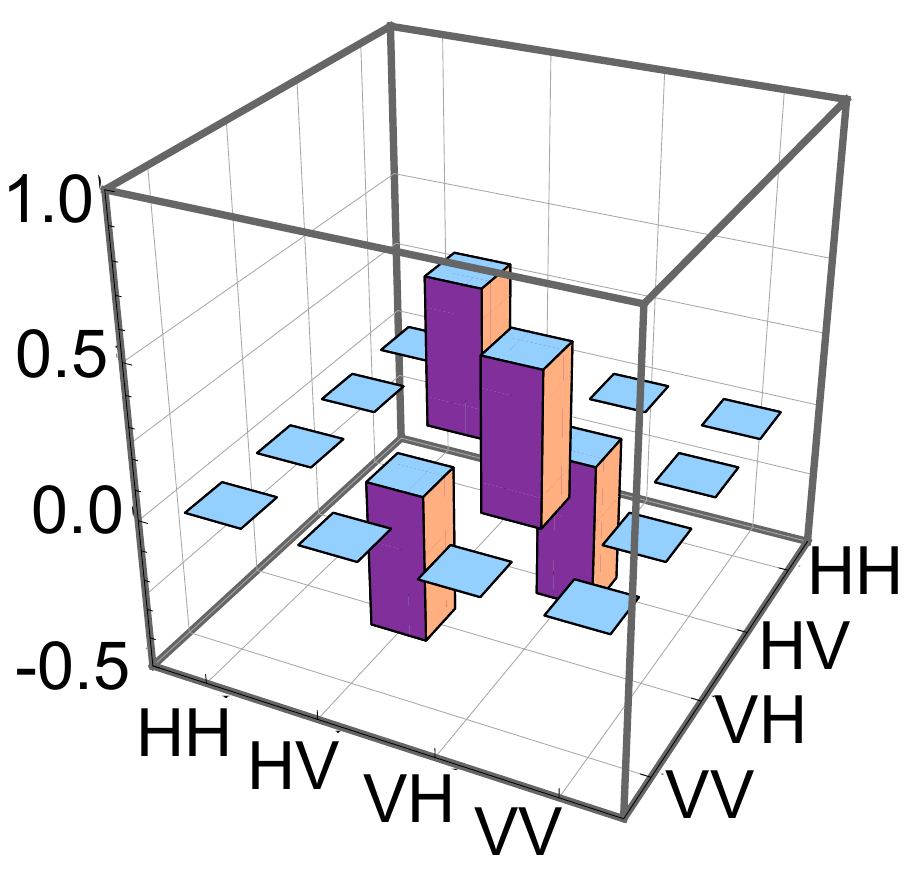}
	\label{fig:densitymatrices_phi-_2_theory}
	\caption{Theoretical density matrix corresponding to the output state: $\ket{\psi_3}=(\ket{01}-\ket{10})/\sqrt{2}$ for all values of $w$ (real parts).}
\end{figure}

\begin{figure}[htbp]
	\centering
		\includegraphics[width=0.50\textwidth]{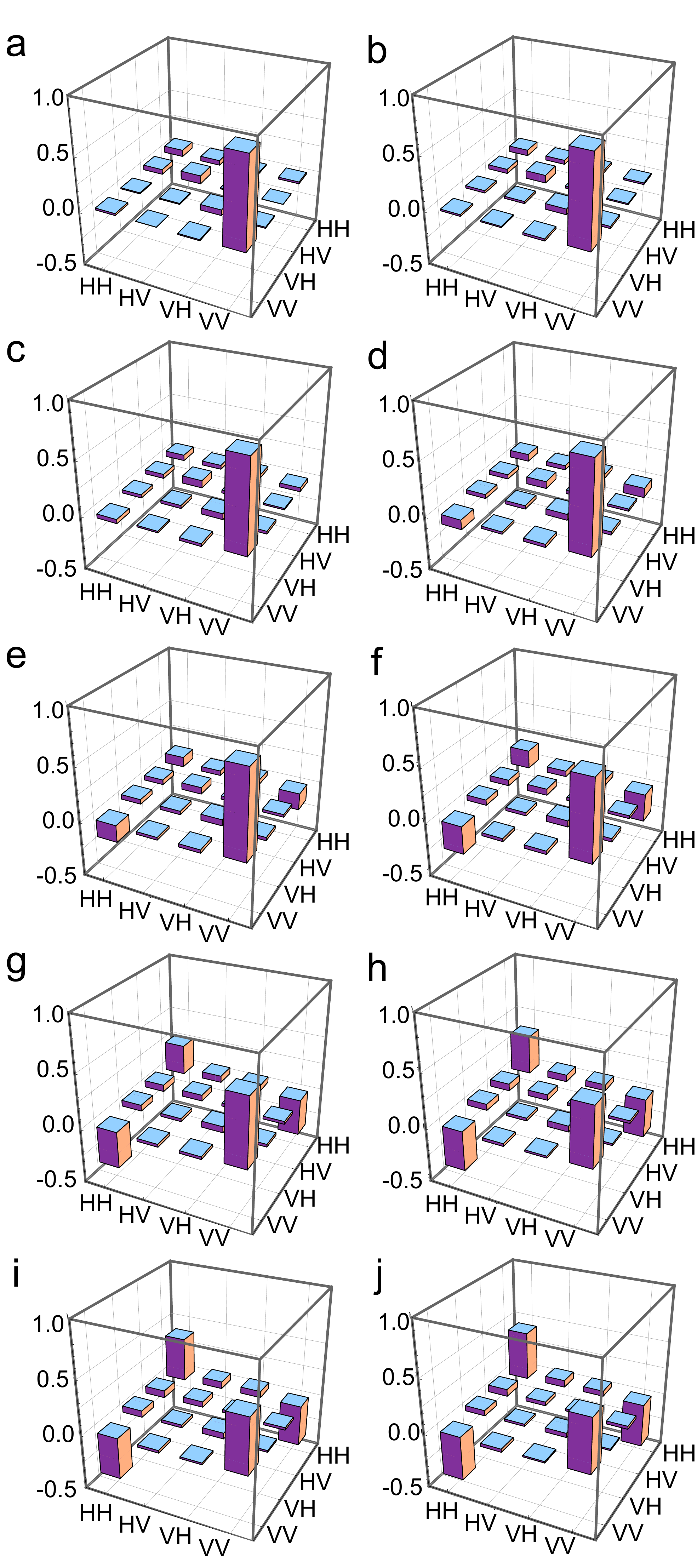}
	\label{fig:densitymatrices_psi-_4}
			\caption{Measured density matrices corresponding to the output state: $\ket{\psi_4}=-\sin{(w/2)}\ket{00}+\cos{(w/2)}\ket{11}$ for different values of $w$ (real parts). a, $w=\arctan{(0)}$, b, $w=\arctan{(1/16)}$, c, $w=\arctan{(1/8)}$, d, $w=\arctan{(1/4)}$, e, $w=\arctan{(1/2)}$, f, $w=\arctan{(1)}$, g, $w=\arctan{(2)}$, h, $w=\arctan{(4)}$, i, $w=\arctan{(4)}$, j, $w=\arctan{(16)}$.}
\end{figure}

\begin{figure}[htbp]
	\centering
		\includegraphics[width=0.50\textwidth]{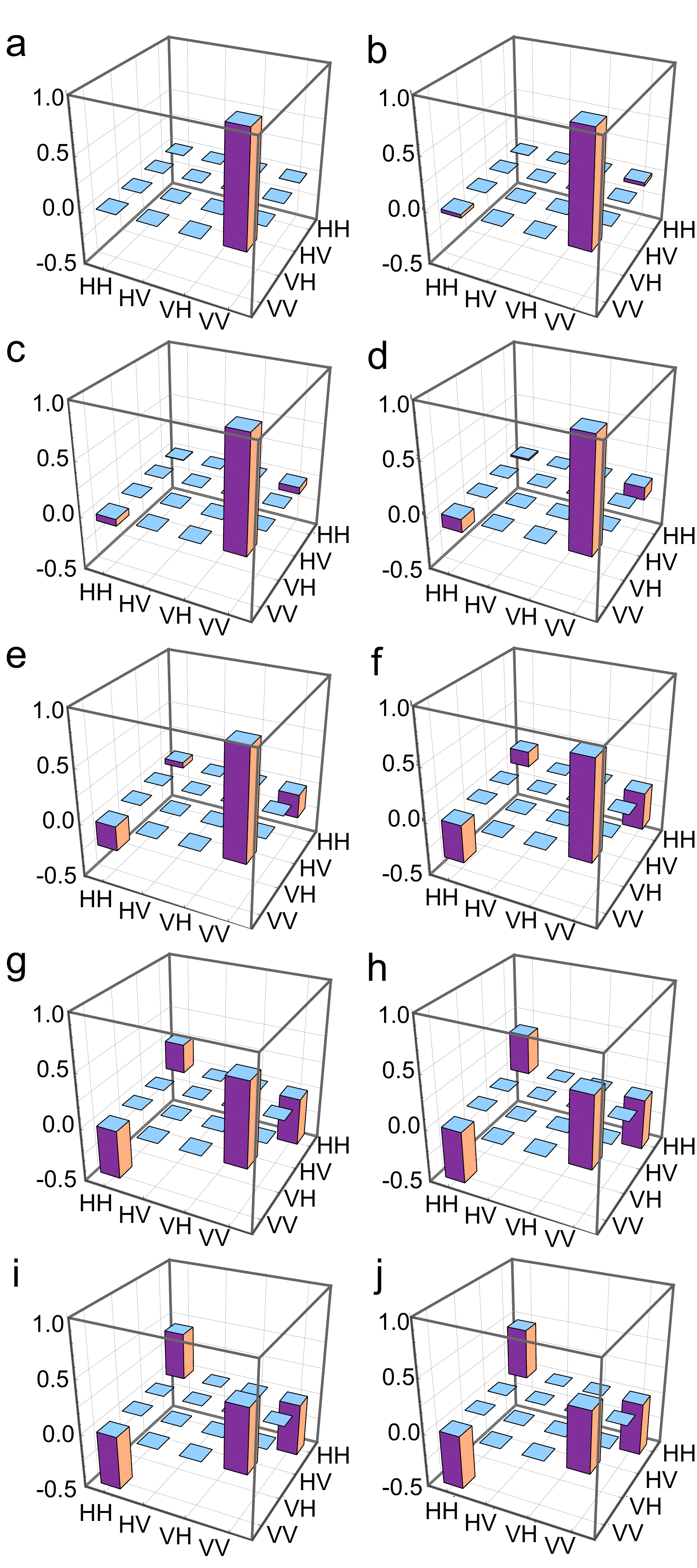}
	\label{fig:densitymatrices_psi-_4_theory}
			\caption{Theoretical density matrices corresponding to the output state: $\ket{\psi_4}=-\sin{(w/2)}\ket{00}+\cos{(w/2)}\ket{11}$ for different values of $w$ (real parts). a, $w=\arctan{(0)}$, b, $w=\arctan{(1/16)}$, c, $w=\arctan{(1/8)}$, d, $w=\arctan{(1/4)}$, e, $w=\arctan{(1/2)}$, f, $w=\arctan{(1)}$, g, $w=\arctan{(2)}$, h, $w=\arctan{(4)}$, i, $w=\arctan{(4)}$, j, $w=\arctan{(16)}$.}
\end{figure}


\begin{thebibliography}{32}%
\makeatletter
\providecommand \@ifxundefined [1]{%
 \@ifx{#1\undefined}
}%
\providecommand \@ifnum [1]{%
 \ifnum #1\expandafter \@firstoftwo
 \else \expandafter \@secondoftwo
 \fi
}%
\providecommand \@ifx [1]{%
 \ifx #1\expandafter \@firstoftwo
 \else \expandafter \@secondoftwo
 \fi
}%
\providecommand \natexlab [1]{#1}%
\providecommand \enquote  [1]{``#1''}%
\providecommand \bibnamefont  [1]{#1}%
\providecommand \bibfnamefont [1]{#1}%
\providecommand \citenamefont [1]{#1}%
\providecommand \href@noop [0]{\@secondoftwo}%
\providecommand \href [0]{\begingroup \@sanitize@url \@href}%
\providecommand \@href[1]{\@@startlink{#1}\@@href}%
\providecommand \@@href[1]{\endgroup#1\@@endlink}%
\providecommand \@sanitize@url [0]{\catcode `\\12\catcode `\$12\catcode
  `\&12\catcode `\#12\catcode `\^12\catcode `\_12\catcode `\%12\relax}%
\providecommand \@@startlink[1]{}%
\providecommand \@@endlink[0]{}%
\providecommand \url  [0]{\begingroup\@sanitize@url \@url }%
\providecommand \@url [1]{\endgroup\@href {#1}{\urlprefix }}%
\providecommand \urlprefix  [0]{URL }%
\providecommand \Eprint [0]{\href }%
\providecommand \doibase [0]{http://dx.doi.org/}%
\providecommand \selectlanguage [0]{\@gobble}%
\providecommand \bibinfo  [0]{\@secondoftwo}%
\providecommand \bibfield  [0]{\@secondoftwo}%
\providecommand \translation [1]{[#1]}%
\providecommand \BibitemOpen [0]{}%
\providecommand \bibitemStop [0]{}%
\providecommand \bibitemNoStop [0]{.\EOS\space}%
\providecommand \EOS [0]{\spacefactor3000\relax}%
\providecommand \BibitemShut  [1]{\csname bibitem#1\endcsname}%
\let\auto@bib@innerbib\@empty
\bibitem [{\citenamefont {Feynman}(1982)}]{feyquantumsimulators}%
  \BibitemOpen
  \bibfield  {author} {\bibinfo {author} {\bibfnamefont {R.}~\bibnamefont
  {Feynman}},\ }\href@noop {} {\bibfield  {journal} {\bibinfo  {journal} {Int.
  J. Theor. Phys.}\ }\textbf {\bibinfo {volume} {21}},\ \bibinfo {pages} {467}
  (\bibinfo {year} {1982})}\BibitemShut {NoStop}%
\bibitem [{\citenamefont {Buluta}\ and\ \citenamefont {Nori}(2009)}]{qsreview}%
  \BibitemOpen
  \bibfield  {author} {\bibinfo {author} {\bibfnamefont {I.}~\bibnamefont
  {Buluta}}\ and\ \bibinfo {author} {\bibfnamefont {F.}~\bibnamefont {Nori}},\
  }\href@noop {} {\bibfield  {journal} {\bibinfo  {journal} {Science}\ }\textbf
  {\bibinfo {volume} {326}},\ \bibinfo {pages} {108} (\bibinfo {year}
  {2009})}\BibitemShut {NoStop}%
\bibitem [{\citenamefont {Aspuru-Guzik}\ and\ \citenamefont
  {Walther}(2012)}]{pqsreview}%
  \BibitemOpen
  \bibfield  {author} {\bibinfo {author} {\bibfnamefont {A.}~\bibnamefont
  {Aspuru-Guzik}}\ and\ \bibinfo {author} {\bibfnamefont {P.}~\bibnamefont
  {Walther}},\ }\href@noop {} {\bibfield  {journal} {\bibinfo  {journal}
  {Nature Phys.}\ }\textbf {\bibinfo {volume} {8}},\ \bibinfo {pages} {285}
  (\bibinfo {year} {2012})}\BibitemShut {NoStop}%
\bibitem [{\citenamefont {Lloyd}(1996)}]{lloyduqs}%
  \BibitemOpen
  \bibfield  {author} {\bibinfo {author} {\bibfnamefont {S.}~\bibnamefont
  {Lloyd}},\ }\href@noop {} {\bibfield  {journal} {\bibinfo  {journal}
  {Science}\ }\textbf {\bibinfo {volume} {273}},\ \bibinfo {pages} {1073}
  (\bibinfo {year} {1996})}\BibitemShut {NoStop}%
\bibitem [{\citenamefont {Abrams}\ and\ \citenamefont
  {Lloyd}(1997)}]{abrcircuitmodel}%
  \BibitemOpen
  \bibfield  {author} {\bibinfo {author} {\bibfnamefont {D.~S.}\ \bibnamefont
  {Abrams}}\ and\ \bibinfo {author} {\bibfnamefont {S.}~\bibnamefont {Lloyd}},\
  }\href@noop {} {\bibfield  {journal} {\bibinfo  {journal} {Phys. Rev. Lett.}\
  }\textbf {\bibinfo {volume} {79}},\ \bibinfo {pages} {2586} (\bibinfo {year}
  {1997})}\BibitemShut {NoStop}%
\bibitem [{\citenamefont {Aspuru-Guzik}\ \emph {et~al.}(2005)\citenamefont
  {Aspuru-Guzik}, \citenamefont {Dutoi}, \citenamefont {Love},\ and\
  \citenamefont {Head-Gordon}}]{guzmolen}%
  \BibitemOpen
  \bibfield  {author} {\bibinfo {author} {\bibfnamefont {A.}~\bibnamefont
  {Aspuru-Guzik}}, \bibinfo {author} {\bibfnamefont {A.~D.}\ \bibnamefont
  {Dutoi}}, \bibinfo {author} {\bibfnamefont {P.~J.}\ \bibnamefont {Love}}, \
  and\ \bibinfo {author} {\bibfnamefont {M.}~\bibnamefont {Head-Gordon}},\
  }\href@noop {} {\bibfield  {journal} {\bibinfo  {journal} {Science}\ }\textbf
  {\bibinfo {volume} {309}},\ \bibinfo {pages} {1704} (\bibinfo {year}
  {2005})}\BibitemShut {NoStop}%
\bibitem [{\citenamefont {Friedenauer}\ \emph {et~al.}(2012)\citenamefont
  {Friedenauer}, \citenamefont {Schmitz}, \citenamefont {Glueckert},
  \citenamefont {Porras},\ and\ \citenamefont {Schaetz}}]{friedenauerions}%
  \BibitemOpen
  \bibfield  {author} {\bibinfo {author} {\bibfnamefont {A.}~\bibnamefont
  {Friedenauer}}, \bibinfo {author} {\bibfnamefont {H.}~\bibnamefont
  {Schmitz}}, \bibinfo {author} {\bibfnamefont {J.~T.}\ \bibnamefont
  {Glueckert}}, \bibinfo {author} {\bibfnamefont {D.}~\bibnamefont {Porras}}, \
  and\ \bibinfo {author} {\bibfnamefont {T.}~\bibnamefont {Schaetz}},\
  }\href@noop {} {\bibfield  {journal} {\bibinfo  {journal} {Nature}\ }\textbf
  {\bibinfo {volume} {484}},\ \bibinfo {pages} {489} (\bibinfo {year}
  {2012})}\BibitemShut {NoStop}%
\bibitem [{\citenamefont {Lanyon}\ \emph {et~al.}(2011)\citenamefont {Lanyon},
  \citenamefont {Hempel}, \citenamefont {Nigg}, \citenamefont {MŸller},
  \citenamefont {Gerritsma}, \citenamefont {ZŠhringer}, \citenamefont
  {Schindler}, \citenamefont {Barreiro}, \citenamefont {Rambach}, \citenamefont
  {Kirchmair}, \citenamefont {Hennrich}, \citenamefont {Zoller}, \citenamefont
  {Blatt},\ and\ \citenamefont {Roos}}]{lanunidqsions}%
  \BibitemOpen
  \bibfield  {author} {\bibinfo {author} {\bibfnamefont {B.}~\bibnamefont
  {Lanyon}}, \bibinfo {author} {\bibfnamefont {C.}~\bibnamefont {Hempel}},
  \bibinfo {author} {\bibfnamefont {D.}~\bibnamefont {Nigg}}, \bibinfo {author}
  {\bibfnamefont {M.}~\bibnamefont {MŸller}}, \bibinfo {author} {\bibfnamefont
  {R.}~\bibnamefont {Gerritsma}}, \bibinfo {author} {\bibfnamefont
  {F.}~\bibnamefont {ZŠhringer}}, \bibinfo {author} {\bibfnamefont
  {P.}~\bibnamefont {Schindler}}, \bibinfo {author} {\bibfnamefont {J.~T.}\
  \bibnamefont {Barreiro}}, \bibinfo {author} {\bibfnamefont {M.}~\bibnamefont
  {Rambach}}, \bibinfo {author} {\bibfnamefont {G.}~\bibnamefont {Kirchmair}},
  \bibinfo {author} {\bibfnamefont {M.}~\bibnamefont {Hennrich}}, \bibinfo
  {author} {\bibfnamefont {P.}~\bibnamefont {Zoller}}, \bibinfo {author}
  {\bibfnamefont {R.}~\bibnamefont {Blatt}}, \ and\ \bibinfo {author}
  {\bibfnamefont {C.~F.}\ \bibnamefont {Roos}},\ }\href@noop {} {\bibfield
  {journal} {\bibinfo  {journal} {Science}\ }\textbf {\bibinfo {volume}
  {334}},\ \bibinfo {pages} {57} (\bibinfo {year} {2011})}\BibitemShut
  {NoStop}%
\bibitem [{\citenamefont {Britton}\ \emph {et~al.}(2012)\citenamefont
  {Britton}, \citenamefont {Sawyer}, \citenamefont {Keith}, \citenamefont
  {Wang}, \citenamefont {Freericks}, \citenamefont {Uys}, \citenamefont
  {Biercuk},\ and\ \citenamefont {Bollinger}}]{brittonions}%
  \BibitemOpen
  \bibfield  {author} {\bibinfo {author} {\bibfnamefont {J.~W.}\ \bibnamefont
  {Britton}}, \bibinfo {author} {\bibfnamefont {B.~C.}\ \bibnamefont {Sawyer}},
  \bibinfo {author} {\bibfnamefont {A.~C.}\ \bibnamefont {Keith}}, \bibinfo
  {author} {\bibfnamefont {C.-C.~J.}\ \bibnamefont {Wang}}, \bibinfo {author}
  {\bibfnamefont {J.~K.}\ \bibnamefont {Freericks}}, \bibinfo {author}
  {\bibfnamefont {H.}~\bibnamefont {Uys}}, \bibinfo {author} {\bibfnamefont
  {M.~J.}\ \bibnamefont {Biercuk}}, \ and\ \bibinfo {author} {\bibfnamefont
  {J.~J.}\ \bibnamefont {Bollinger}},\ }\href@noop {} {\bibfield  {journal}
  {\bibinfo  {journal} {Nature}\ }\textbf {\bibinfo {volume} {484}},\ \bibinfo
  {pages} {489} (\bibinfo {year} {2012})}\BibitemShut {NoStop}%
\bibitem [{\citenamefont {Kim}\ \emph {et~al.}(2010)\citenamefont {Kim},
  \citenamefont {Chang}, \citenamefont {Korenblit}, \citenamefont {Islam},
  \citenamefont {Edwards}, \citenamefont {Freericks}, \citenamefont {Lin},
  \citenamefont {Duan},\ and\ \citenamefont {Monroe}}]{kimions}%
  \BibitemOpen
  \bibfield  {author} {\bibinfo {author} {\bibfnamefont {K.}~\bibnamefont
  {Kim}}, \bibinfo {author} {\bibfnamefont {M.-S.}\ \bibnamefont {Chang}},
  \bibinfo {author} {\bibfnamefont {S.}~\bibnamefont {Korenblit}}, \bibinfo
  {author} {\bibfnamefont {R.}~\bibnamefont {Islam}}, \bibinfo {author}
  {\bibfnamefont {E.~E.}\ \bibnamefont {Edwards}}, \bibinfo {author}
  {\bibfnamefont {J.~K.}\ \bibnamefont {Freericks}}, \bibinfo {author}
  {\bibfnamefont {G.-D.}\ \bibnamefont {Lin}}, \bibinfo {author} {\bibfnamefont
  {L.-M.}\ \bibnamefont {Duan}}, \ and\ \bibinfo {author} {\bibfnamefont
  {C.}~\bibnamefont {Monroe}},\ }\href@noop {} {\bibfield  {journal} {\bibinfo
  {journal} {Nature}\ }\textbf {\bibinfo {volume} {465}},\ \bibinfo {pages}
  {590} (\bibinfo {year} {2010})}\BibitemShut {NoStop}%
\bibitem [{\citenamefont {Trotzky}\ \emph {et~al.}(2008)\citenamefont
  {Trotzky}, \citenamefont {Cheinet}, \citenamefont {Fšlling}, \citenamefont
  {Feld}, \citenamefont {Schnorrberger}, \citenamefont {Rey}, \citenamefont
  {Polkovnikov}, \citenamefont {Demler}, \citenamefont {Lukin},\ and\
  \citenamefont {Bloch}}]{trotzkyol}%
  \BibitemOpen
  \bibfield  {author} {\bibinfo {author} {\bibfnamefont {S.}~\bibnamefont
  {Trotzky}}, \bibinfo {author} {\bibfnamefont {P.}~\bibnamefont {Cheinet}},
  \bibinfo {author} {\bibfnamefont {S.}~\bibnamefont {Fšlling}}, \bibinfo
  {author} {\bibfnamefont {M.}~\bibnamefont {Feld}}, \bibinfo {author}
  {\bibfnamefont {U.}~\bibnamefont {Schnorrberger}}, \bibinfo {author}
  {\bibfnamefont {A.~M.}\ \bibnamefont {Rey}}, \bibinfo {author} {\bibfnamefont
  {A.}~\bibnamefont {Polkovnikov}}, \bibinfo {author} {\bibfnamefont {E.~A.}\
  \bibnamefont {Demler}}, \bibinfo {author} {\bibfnamefont {M.~D.}\
  \bibnamefont {Lukin}}, \ and\ \bibinfo {author} {\bibfnamefont
  {I.}~\bibnamefont {Bloch}},\ }\href@noop {} {\bibfield  {journal} {\bibinfo
  {journal} {Science}\ }\textbf {\bibinfo {volume} {319}},\ \bibinfo {pages}
  {295} (\bibinfo {year} {2008})}\BibitemShut {NoStop}%
\bibitem [{\citenamefont {Simon}\ \emph {et~al.}(2011)\citenamefont {Simon},
  \citenamefont {Bakr}, \citenamefont {Ma}, \citenamefont {Tai}, \citenamefont
  {Preiss},\ and\ \citenamefont {Greiner}}]{simonol}%
  \BibitemOpen
  \bibfield  {author} {\bibinfo {author} {\bibfnamefont {J.}~\bibnamefont
  {Simon}}, \bibinfo {author} {\bibfnamefont {W.~S.}\ \bibnamefont {Bakr}},
  \bibinfo {author} {\bibfnamefont {R.}~\bibnamefont {Ma}}, \bibinfo {author}
  {\bibfnamefont {M.~E.}\ \bibnamefont {Tai}}, \bibinfo {author} {\bibfnamefont
  {P.~M.}\ \bibnamefont {Preiss}}, \ and\ \bibinfo {author} {\bibfnamefont
  {M.}~\bibnamefont {Greiner}},\ }\href@noop {} {\bibfield  {journal} {\bibinfo
   {journal} {Nature}\ }\textbf {\bibinfo {volume} {472}},\ \bibinfo {pages}
  {307} (\bibinfo {year} {2011})}\BibitemShut {NoStop}%
\bibitem [{\citenamefont {Somaroo}\ \emph {et~al.}(1999)\citenamefont
  {Somaroo}, \citenamefont {Tseng}, \citenamefont {Havel}, \citenamefont
  {Laflamme},\ and\ \citenamefont {Cory}}]{somaroonmr}%
  \BibitemOpen
  \bibfield  {author} {\bibinfo {author} {\bibfnamefont {S.}~\bibnamefont
  {Somaroo}}, \bibinfo {author} {\bibfnamefont {C.~H.}\ \bibnamefont {Tseng}},
  \bibinfo {author} {\bibfnamefont {T.~F.}\ \bibnamefont {Havel}}, \bibinfo
  {author} {\bibfnamefont {R.}~\bibnamefont {Laflamme}}, \ and\ \bibinfo
  {author} {\bibfnamefont {D.~G.}\ \bibnamefont {Cory}},\ }\href@noop {}
  {\bibfield  {journal} {\bibinfo  {journal} {Phys. Rev. Lett.}\ }\textbf
  {\bibinfo {volume} {82}},\ \bibinfo {pages} {5381} (\bibinfo {year}
  {1999})}\BibitemShut {NoStop}%
\bibitem [{\citenamefont {Negrevergne}\ \emph {et~al.}(2005)\citenamefont
  {Negrevergne}, \citenamefont {Somma}, \citenamefont {Ortiz}, \citenamefont
  {Knill},\ and\ \citenamefont {Laflamme}}]{negrevergnenmr}%
  \BibitemOpen
  \bibfield  {author} {\bibinfo {author} {\bibfnamefont {C.}~\bibnamefont
  {Negrevergne}}, \bibinfo {author} {\bibfnamefont {R.}~\bibnamefont {Somma}},
  \bibinfo {author} {\bibfnamefont {G.}~\bibnamefont {Ortiz}}, \bibinfo
  {author} {\bibfnamefont {E.}~\bibnamefont {Knill}}, \ and\ \bibinfo {author}
  {\bibfnamefont {R.}~\bibnamefont {Laflamme}},\ }\href@noop {} {\bibfield
  {journal} {\bibinfo  {journal} {Phys. Rev. A}\ }\textbf {\bibinfo {volume}
  {71}},\ \bibinfo {pages} {032344} (\bibinfo {year} {2005})}\BibitemShut
  {NoStop}%
\bibitem [{\citenamefont {Brown}\ \emph {et~al.}(2006)\citenamefont {Brown},
  \citenamefont {Clark},\ and\ \citenamefont {Chuang}}]{brownnmr}%
  \BibitemOpen
  \bibfield  {author} {\bibinfo {author} {\bibfnamefont {K.~R.}\ \bibnamefont
  {Brown}}, \bibinfo {author} {\bibfnamefont {R.~J.}\ \bibnamefont {Clark}}, \
  and\ \bibinfo {author} {\bibfnamefont {I.~L.}\ \bibnamefont {Chuang}},\
  }\href@noop {} {\bibfield  {journal} {\bibinfo  {journal} {Phys. Rev. Lett.}\
  }\textbf {\bibinfo {volume} {97}},\ \bibinfo {pages} {050504} (\bibinfo
  {year} {2006})}\BibitemShut {NoStop}%
\bibitem [{\citenamefont {Du}\ \emph {et~al.}(2010)\citenamefont {Du},
  \citenamefont {Xu}, \citenamefont {Peng}, \citenamefont {Wang}, \citenamefont
  {Wu},\ and\ \citenamefont {Lu}}]{dunmr}%
  \BibitemOpen
  \bibfield  {author} {\bibinfo {author} {\bibfnamefont {J.}~\bibnamefont
  {Du}}, \bibinfo {author} {\bibfnamefont {N.}~\bibnamefont {Xu}}, \bibinfo
  {author} {\bibfnamefont {X.}~\bibnamefont {Peng}}, \bibinfo {author}
  {\bibfnamefont {P.}~\bibnamefont {Wang}}, \bibinfo {author} {\bibfnamefont
  {S.}~\bibnamefont {Wu}}, \ and\ \bibinfo {author} {\bibfnamefont
  {D.}~\bibnamefont {Lu}},\ }\href@noop {} {\bibfield  {journal} {\bibinfo
  {journal} {Phys. Rev. Lett.}\ }\textbf {\bibinfo {volume} {104}},\ \bibinfo
  {pages} {030502} (\bibinfo {year} {2010})}\BibitemShut {NoStop}%
\bibitem [{\citenamefont {Peng}\ \emph {et~al.}(2010)\citenamefont {Peng},
  \citenamefont {Wu}, \citenamefont {Li}, \citenamefont {Suter},\ and\
  \citenamefont {Du}}]{Peng2010}%
  \BibitemOpen
  \bibfield  {author} {\bibinfo {author} {\bibfnamefont {X.}~\bibnamefont
  {Peng}}, \bibinfo {author} {\bibfnamefont {S.}~\bibnamefont {Wu}}, \bibinfo
  {author} {\bibfnamefont {J.}~\bibnamefont {Li}}, \bibinfo {author}
  {\bibfnamefont {D.}~\bibnamefont {Suter}}, \ and\ \bibinfo {author}
  {\bibfnamefont {J.}~\bibnamefont {Du}},\ }\href@noop {} {\bibfield  {journal}
  {\bibinfo  {journal} {Phys. Rev. Lett.}\ }\textbf {\bibinfo {volume} {105}},\
  \bibinfo {pages} {240405} (\bibinfo {year} {2010})}\BibitemShut {NoStop}%
\bibitem [{\citenamefont {Kassal}\ \emph {et~al.}(2008)\citenamefont {Kassal},
  \citenamefont {Jordan}, \citenamefont {Love}, \citenamefont {Mohsenia},\ and\
  \citenamefont {Aspuru-Guzik}}]{kassalph}%
  \BibitemOpen
  \bibfield  {author} {\bibinfo {author} {\bibfnamefont {I.}~\bibnamefont
  {Kassal}}, \bibinfo {author} {\bibfnamefont {S.~P.}\ \bibnamefont {Jordan}},
  \bibinfo {author} {\bibfnamefont {P.~J.}\ \bibnamefont {Love}}, \bibinfo
  {author} {\bibfnamefont {M.}~\bibnamefont {Mohsenia}}, \ and\ \bibinfo
  {author} {\bibfnamefont {A.}~\bibnamefont {Aspuru-Guzik}},\ }\href@noop {}
  {\bibfield  {journal} {\bibinfo  {journal} {Proc. Nat. Acad. Sci. USA}\
  }\textbf {\bibinfo {volume} {105}},\ \bibinfo {pages} {18681} (\bibinfo
  {year} {2008})}\BibitemShut {NoStop}%
\bibitem [{\citenamefont {Ma}\ \emph {et~al.}(2011)\citenamefont {Ma},
  \citenamefont {Daki?}, \citenamefont {Naylor}, \citenamefont {Zeilinger},\
  and\ \citenamefont {Walther}}]{mafrust}%
  \BibitemOpen
  \bibfield  {author} {\bibinfo {author} {\bibfnamefont {X.-S.}\ \bibnamefont
  {Ma}}, \bibinfo {author} {\bibfnamefont {B.}~\bibnamefont {Daki?}}, \bibinfo
  {author} {\bibfnamefont {W.}~\bibnamefont {Naylor}}, \bibinfo {author}
  {\bibfnamefont {A.}~\bibnamefont {Zeilinger}}, \ and\ \bibinfo {author}
  {\bibfnamefont {P.}~\bibnamefont {Walther}},\ }\href@noop {} {\bibfield
  {journal} {\bibinfo  {journal} {Nature Phys.}\ }\textbf {\bibinfo {volume}
  {7}},\ \bibinfo {pages} {399} (\bibinfo {year} {2011})}\BibitemShut {NoStop}%
\bibitem [{\citenamefont {Orieux}\ \emph {et~al.}(2013)\citenamefont {Orieux},
  \citenamefont {Boutari}, \citenamefont {Barbieri}, \citenamefont
  {Paternostro},\ and\ \citenamefont {Mataloni}}]{Orieux2013}%
  \BibitemOpen
  \bibfield  {author} {\bibinfo {author} {\bibfnamefont {A.}~\bibnamefont
  {Orieux}}, \bibinfo {author} {\bibfnamefont {J.}~\bibnamefont {Boutari}},
  \bibinfo {author} {\bibfnamefont {M.}~\bibnamefont {Barbieri}}, \bibinfo
  {author} {\bibfnamefont {M.}~\bibnamefont {Paternostro}}, \ and\ \bibinfo
  {author} {\bibfnamefont {P.}~\bibnamefont {Mataloni}},\ }\href@noop {}
  {\bibfield  {journal} {\bibinfo  {journal} {arXiv preprint arXiv:1312.1102}\
  } (\bibinfo {year} {2013})}\BibitemShut {NoStop}%
\bibitem [{\citenamefont {Verstraete}\ \emph {et~al.}(2009)\citenamefont
  {Verstraete}, \citenamefont {Cirac},\ and\ \citenamefont
  {Latorre}}]{Verstraete2008}%
  \BibitemOpen
  \bibfield  {author} {\bibinfo {author} {\bibfnamefont {F.}~\bibnamefont
  {Verstraete}}, \bibinfo {author} {\bibfnamefont {J.~I.}\ \bibnamefont
  {Cirac}}, \ and\ \bibinfo {author} {\bibfnamefont {J.~I.}\ \bibnamefont
  {Latorre}},\ }\href@noop {} {\bibfield  {journal} {\bibinfo  {journal} {Phys.
  Rev. A}\ }\textbf {\bibinfo {volume} {79}},\ \bibinfo {pages} {032316}
  (\bibinfo {year} {2009})}\BibitemShut {NoStop}%
\bibitem [{\citenamefont {Kwiat}\ \emph {et~al.}(1999)\citenamefont {Kwiat},
  \citenamefont {Waks}, \citenamefont {White}, \citenamefont {Appelbaum},\ and\
  \citenamefont {Eberhard}}]{Kwiat1999}%
  \BibitemOpen
  \bibfield  {author} {\bibinfo {author} {\bibfnamefont {P.~G.}\ \bibnamefont
  {Kwiat}}, \bibinfo {author} {\bibfnamefont {E.}~\bibnamefont {Waks}},
  \bibinfo {author} {\bibfnamefont {A.~G.}\ \bibnamefont {White}}, \bibinfo
  {author} {\bibfnamefont {I.}~\bibnamefont {Appelbaum}}, \ and\ \bibinfo
  {author} {\bibfnamefont {P.~H.}\ \bibnamefont {Eberhard}},\ }\href {\doibase
  10.1103/PhysRevA.60.R773} {\bibfield  {journal} {\bibinfo  {journal} {Phys.
  Rev. A}\ }\textbf {\bibinfo {volume} {60}},\ \bibinfo {pages} {R773}
  (\bibinfo {year} {1999})}\BibitemShut {NoStop}%
\bibitem [{\citenamefont {Langford}\ \emph {et~al.}(2005)\citenamefont
  {Langford}, \citenamefont {Weinhold}, \citenamefont {Prevedel}, \citenamefont
  {Resch}, \citenamefont {Gilchrist}, \citenamefont {O'Brien}, \citenamefont
  {Pryde},\ and\ \citenamefont {White}}]{lancsign}%
  \BibitemOpen
  \bibfield  {author} {\bibinfo {author} {\bibfnamefont {N.~K.}\ \bibnamefont
  {Langford}}, \bibinfo {author} {\bibfnamefont {T.~J.}\ \bibnamefont
  {Weinhold}}, \bibinfo {author} {\bibfnamefont {R.}~\bibnamefont {Prevedel}},
  \bibinfo {author} {\bibfnamefont {K.~J.}\ \bibnamefont {Resch}}, \bibinfo
  {author} {\bibfnamefont {A.}~\bibnamefont {Gilchrist}}, \bibinfo {author}
  {\bibfnamefont {J.~L.}\ \bibnamefont {O'Brien}}, \bibinfo {author}
  {\bibfnamefont {G.~J.}\ \bibnamefont {Pryde}}, \ and\ \bibinfo {author}
  {\bibfnamefont {A.~G.}\ \bibnamefont {White}},\ }\href@noop {} {\bibfield
  {journal} {\bibinfo  {journal} {Phys. Rev. Lett.}\ }\textbf {\bibinfo
  {volume} {95}},\ \bibinfo {pages} {210504} (\bibinfo {year}
  {2005})}\BibitemShut {NoStop}%
\bibitem [{\citenamefont {Kiesel}\ \emph
  {et~al.}(2005{\natexlab{a}})\citenamefont {Kiesel}, \citenamefont {Schmid},
  \citenamefont {Weber}, \citenamefont {Ursin},\ and\ \citenamefont
  {Weinfurter}}]{kiecsign}%
  \BibitemOpen
  \bibfield  {author} {\bibinfo {author} {\bibfnamefont {N.}~\bibnamefont
  {Kiesel}}, \bibinfo {author} {\bibfnamefont {C.}~\bibnamefont {Schmid}},
  \bibinfo {author} {\bibfnamefont {U.}~\bibnamefont {Weber}}, \bibinfo
  {author} {\bibfnamefont {R.}~\bibnamefont {Ursin}}, \ and\ \bibinfo {author}
  {\bibfnamefont {H.}~\bibnamefont {Weinfurter}},\ }\href@noop {} {\bibfield
  {journal} {\bibinfo  {journal} {Phys. Rev. Lett.}\ }\textbf {\bibinfo
  {volume} {95}},\ \bibinfo {pages} {210505} (\bibinfo {year}
  {2005}{\natexlab{a}})}\BibitemShut {NoStop}%
\bibitem [{\citenamefont {Okamoto}\ \emph {et~al.}(2005)\citenamefont
  {Okamoto}, \citenamefont {Hofmann}, \citenamefont {Takeuchi},\ and\
  \citenamefont {Sasaki}}]{okacsign}%
  \BibitemOpen
  \bibfield  {author} {\bibinfo {author} {\bibfnamefont {R.}~\bibnamefont
  {Okamoto}}, \bibinfo {author} {\bibfnamefont {H.~F.}\ \bibnamefont
  {Hofmann}}, \bibinfo {author} {\bibfnamefont {S.}~\bibnamefont {Takeuchi}}, \
  and\ \bibinfo {author} {\bibfnamefont {K.}~\bibnamefont {Sasaki}},\
  }\href@noop {} {\bibfield  {journal} {\bibinfo  {journal} {Phys. Rev. Lett.}\
  }\textbf {\bibinfo {volume} {95}},\ \bibinfo {pages} {210506} (\bibinfo
  {year} {2005})}\BibitemShut {NoStop}%
\bibitem [{\citenamefont {Kiesel}\ \emph
  {et~al.}(2005{\natexlab{b}})\citenamefont {Kiesel}, \citenamefont {Schmid},
  \citenamefont {Weber}, \citenamefont {Ursin},\ and\ \citenamefont
  {Weinfurter}}]{Kiesel2005a}%
  \BibitemOpen
  \bibfield  {author} {\bibinfo {author} {\bibfnamefont {N.}~\bibnamefont
  {Kiesel}}, \bibinfo {author} {\bibfnamefont {C.}~\bibnamefont {Schmid}},
  \bibinfo {author} {\bibfnamefont {U.}~\bibnamefont {Weber}}, \bibinfo
  {author} {\bibfnamefont {R.}~\bibnamefont {Ursin}}, \ and\ \bibinfo {author}
  {\bibfnamefont {H.}~\bibnamefont {Weinfurter}},\ }\href {\doibase
  10.1103/PhysRevLett.95.210505} {\bibfield  {journal} {\bibinfo  {journal}
  {Phys. Rev. Lett.}\ }\textbf {\bibinfo {volume} {95}},\ \bibinfo {pages}
  {210505} (\bibinfo {year} {2005}{\natexlab{b}})}\BibitemShut {NoStop}%
\bibitem [{\citenamefont {White}\ \emph {et~al.}(2007)\citenamefont {White},
  \citenamefont {Gilchrist}, \citenamefont {Pryde}, \citenamefont {O'Brien},
  \citenamefont {Bremner},\ and\ \citenamefont {Langford}}]{whimeasuringgates}%
  \BibitemOpen
  \bibfield  {author} {\bibinfo {author} {\bibfnamefont {A.~G.}\ \bibnamefont
  {White}}, \bibinfo {author} {\bibfnamefont {A.}~\bibnamefont {Gilchrist}},
  \bibinfo {author} {\bibfnamefont {G.}~\bibnamefont {Pryde}}, \bibinfo
  {author} {\bibfnamefont {J.~L.}\ \bibnamefont {O'Brien}}, \bibinfo {author}
  {\bibfnamefont {M.}~\bibnamefont {Bremner}}, \ and\ \bibinfo {author}
  {\bibfnamefont {N.~K.}\ \bibnamefont {Langford}},\ }\href@noop {} {\bibfield
  {journal} {\bibinfo  {journal} {J. Opt. Soc. Am. B}\ }\textbf {\bibinfo
  {volume} {24}},\ \bibinfo {pages} {172} (\bibinfo {year} {2007})}\BibitemShut
  {NoStop}%
\bibitem [{\citenamefont {Chen}\ and\ \citenamefont
  {Nussinov}(2008)}]{kitaevanalytical}%
  \BibitemOpen
  \bibfield  {author} {\bibinfo {author} {\bibfnamefont {H.-D.}\ \bibnamefont
  {Chen}}\ and\ \bibinfo {author} {\bibfnamefont {Z.}~\bibnamefont
  {Nussinov}},\ }\href@noop {} {\bibfield  {journal} {\bibinfo  {journal} {J.
  Phys. A: Math. Theor.}\ }\textbf {\bibinfo {volume} {41}},\ \bibinfo {pages}
  {075001} (\bibinfo {year} {2008})}\BibitemShut {NoStop}%
\bibitem [{\citenamefont {Polkovnikov}\ \emph {et~al.}(2011)\citenamefont
  {Polkovnikov}, \citenamefont {Sengupta}, \citenamefont {Silva},\ and\
  \citenamefont {Vengalattore}}]{Polkovnikov2011}%
  \BibitemOpen
  \bibfield  {author} {\bibinfo {author} {\bibfnamefont {A.}~\bibnamefont
  {Polkovnikov}}, \bibinfo {author} {\bibfnamefont {K.}~\bibnamefont
  {Sengupta}}, \bibinfo {author} {\bibfnamefont {A.}~\bibnamefont {Silva}}, \
  and\ \bibinfo {author} {\bibfnamefont {M.}~\bibnamefont {Vengalattore}},\
  }\href {\doibase 10.1103/RevModPhys.83.863} {\bibfield  {journal} {\bibinfo
  {journal} {Rev. Mod. Phys.}\ }\textbf {\bibinfo {volume} {83}},\ \bibinfo
  {pages} {863} (\bibinfo {year} {2011})}\BibitemShut {NoStop}%
\bibitem [{\citenamefont {Cazalilla}\ and\ \citenamefont
  {Rigol}(2010)}]{Cazalilla2010}%
  \BibitemOpen
  \bibfield  {author} {\bibinfo {author} {\bibfnamefont {M.~A.}\ \bibnamefont
  {Cazalilla}}\ and\ \bibinfo {author} {\bibfnamefont {M.}~\bibnamefont
  {Rigol}},\ }\href {\doibase 10.1088/1367-2630/12/5/055006} {\bibfield
  {journal} {\bibinfo  {journal} {New J. Phys.}\ }\textbf {\bibinfo {volume}
  {12}},\ \bibinfo {pages} {055006} (\bibinfo {year} {2010})}\BibitemShut
  {NoStop}%
\bibitem [{\citenamefont {Mart{\'\i}n-L{\'o}pez}\ \emph
  {et~al.}(2012)\citenamefont {Mart{\'\i}n-L{\'o}pez}, \citenamefont {Laing},
  \citenamefont {Lawson}, \citenamefont {Alvarez}, \citenamefont {Zhou},\ and\
  \citenamefont {O'Brien}}]{lopez2012}%
  \BibitemOpen
  \bibfield  {author} {\bibinfo {author} {\bibfnamefont {E.}~\bibnamefont
  {Mart{\'\i}n-L{\'o}pez}}, \bibinfo {author} {\bibfnamefont {A.}~\bibnamefont
  {Laing}}, \bibinfo {author} {\bibfnamefont {T.}~\bibnamefont {Lawson}},
  \bibinfo {author} {\bibfnamefont {R.}~\bibnamefont {Alvarez}}, \bibinfo
  {author} {\bibfnamefont {X.-Q.}\ \bibnamefont {Zhou}}, \ and\ \bibinfo
  {author} {\bibfnamefont {J.~L.}\ \bibnamefont {O'Brien}},\ }\href@noop {}
  {\bibfield  {journal} {\bibinfo  {journal} {Nature Photonics}\ }\textbf
  {\bibinfo {volume} {6}},\ \bibinfo {pages} {773} (\bibinfo {year}
  {2012})}\BibitemShut {NoStop}%
\bibitem [{\citenamefont {Barz}\ \emph {et~al.}(2014)\citenamefont {Barz},
  \citenamefont {Kassal}, \citenamefont {Ringbauer}, \citenamefont {Lipp},
  \citenamefont {B.~Dakic},\ and\ \citenamefont {Walther}}]{Barz2014}%
  \BibitemOpen
  \bibfield  {author} {\bibinfo {author} {\bibfnamefont {S.}~\bibnamefont
  {Barz}}, \bibinfo {author} {\bibfnamefont {I.}~\bibnamefont {Kassal}},
  \bibinfo {author} {\bibfnamefont {M.}~\bibnamefont {Ringbauer}}, \bibinfo
  {author} {\bibfnamefont {Y.~O.}\ \bibnamefont {Lipp}}, \bibinfo {author}
  {\bibfnamefont {A.~A.-G.}\ \bibnamefont {B.~Dakic}}, \ and\ \bibinfo {author}
  {\bibfnamefont {P.}~\bibnamefont {Walther}},\ }\href@noop {} {\bibfield
  {journal} {\bibinfo  {journal} {Scientific Reports}\ }\textbf {\bibinfo
  {volume} {4}},\ \bibinfo {pages} {6115} (\bibinfo {year} {2014})}\BibitemShut
  {NoStop}%
\end{thebibliography}
\end{document}